\documentclass[journal]{IEEEtran}

\usepackage{graphics} 
\usepackage{epsfig} 
\usepackage{times} 
\usepackage{amsmath} 
\usepackage{amssymb}  
\usepackage{epstopdf}
\usepackage{multirow}
\usepackage{color} 

\usepackage{algorithm} 
\usepackage{algpseudocode}
\algrenewcommand\algorithmicindent{1.2em}

\usepackage{cite}
\usepackage{subfig}

\DeclareMathOperator*{\argmin}{arg\!\min}

\renewcommand{\baselinestretch}{0.938}

\title{\LARGE \bf
A Unified Safety Protection and Extension Governor
}
\author{Nan Li, Yutong Li, and Ilya Kolmanovsky
\vspace{-5mm}
\thanks{This research is supported by the Ford Motor Company.}
\thanks{Nan Li is with the Department of Aerospace Engineering, Auburn University, Auburn, AL 36849, USA ({\tt email: nanli@auburn.edu}). Yutong Li is with the Ford Motor Company, Dearborn, MI 48126, USA. Ilya Kolmanovsky is with the Department of Aerospace Engineering, University of Michigan, Ann Arbor, MI 48109, USA.}}

\let\labelindent\relax

\begin{document}

\maketitle
\thispagestyle{empty}
\pagestyle{empty}

\begin{abstract}
In this paper, we propose a supervisory control scheme that unifies the abilities of safety protection and safety extension. It produces a control that is able to keep the system safe indefinitely when such a control exists. When such a control does not exist due to abnormal system states, it optimizes the control to maximize the time before any safety violation, which translates into more time to seek recovery and/or mitigate any harm. We describe the scheme and develop an approach that integrates the two capabilities into a single constrained optimization problem with only continuous variables. For linear systems with convex constraints, the problem reduces to a convex quadratic program and is easy to solve. We illustrate the proposed safety supervisor with an automotive example.
\end{abstract}

\vspace{-2mm}
\section{Introduction}\label{sec:intro}

Many safety-related specifications for autonomous system operation, such as collision avoidance, power or thermal limits, can be expressed as constraints on system state and input variables. Control approaches that can explicitly handle constraints include the model predictive control (MPC)~\cite{mayne2014model,mayne2000constrained}, control Lyapunov/barrier functions~\cite{tee2009barrier,ames2016control,jankovic2018robust}, reference governors~\cite{garone2017reference}, and action governors~\cite{li2021robust}, etc. Many of these approaches use repeated online optimization to determine a control solution that respects the constraints over a prediction window. Therefore, they can keep the system safe if the formulated optimization problem admits such a control solution at each time, called recursively feasible. 

A recursive feasibility guarantee typically relies on assumptions of feasibility at the initial time and bounds of any uncertainties/disturbances~\cite{mayne2014model,jankovic2018robust,garone2017reference,li2021robust}. However, these assumptions may be violated during the actual system operation, for instance, due to sensor/actuator faults or rare disturbance inputs that are not within the assumed bounds. This may cause the online optimization problem to become infeasible. Practical strategies for handling a feasibility loss include relaxing hard constraints to soft constraints~\cite{mayne2000constrained} and keeping the reference input constant in the reference governor approach~\cite{garone2017reference}. However, these strategies typically do not provide a performance guarantee. Compared to the effort put in developing recursive feasibility guarantees (based on certain assumptions) for satisfying constraints, less effort has been made to address, in a systematic manner, the situation of a feasibility loss, or more generally, situations where no solution respecting constraints over the entire prediction window exists. 

The drift counteraction optimal control (DCOC) is an approach to addressing such a situation~\cite{kolmanovsky2008discrete,tang2022continuous}. The goal of DCOC is to delay the first constraint violation as much as possible assuming an eventual constraint violation is unavoidable. In this regard, the DCOC can be viewed as a time-optimal control problem, where we aim to maximize the time before the system reaches a state that is unsafe. In terms of improving safety, this strategy is reasonable because extending the duration before actual safety violation translates into more time to respond to seek recovery or mitigate potential harm. 

Motivated by the aforementioned infeasibility issue and the DCOC strategy, in this paper we develop an approach that achieves the following two goals simultaneously: {\bf 1)~ Safety Protection}: When there is a control that is determined to be able to keep the system safe over the entire prediction window (which is from the present time to infinite time in this paper), it is obtained as the solution. {\bf 2)~ Safety Extension}: When such a control does not exist, the approach computes a control that maximizes the duration of constraint violation-free operation. 

\begin{figure}[thpb]
      \centering
      \includegraphics[width=240pt]{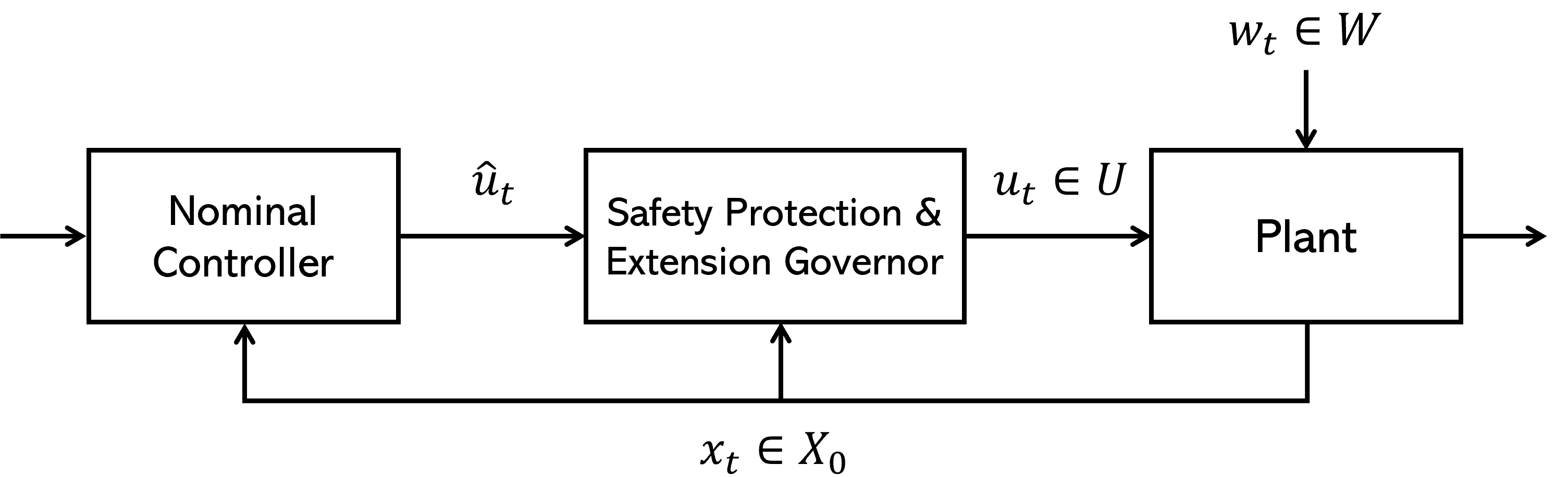}
      \caption{Control system architecture with the proposed safety protection \& extension governor.}
      \label{fig:1}
\end{figure}

In particular, we focus on the design of a supervisory scheme that monitors the control input generated by a nominal controller and corrects it when the nominal control may cause safety violations (see Fig.~\ref{fig:1}). The designed safety supervisor is independent of the nominal controller and allows it to be nonlinear and time-varying. For instance, the nominal controller can be a neural network and/or evolve over time through online learning~\cite{li2021safe}. The proposed scheme acts similarly to the control barrier function-based safety filter~\cite{ames2016control,jankovic2018robust} and the action governor~\cite{li2021robust}. A unique feature of the proposed scheme is its ability to extend safety when a solution determined to guarantee safety indefinitely does not exist (due to the various reasons discussed above). This {\bf Safety Extension} ability and the seamless integration of this ability with the {\bf Safety Protection} ability (as defined above) are the main contributions of this paper. Indeed, as will be shown, the {\bf Safety Protection} ability of the supervisor can be implemented through a barrier function or an action governor, while our approach in this paper will add the {\bf Safety Extension} ability to the supervisor, making it more capable and hence practically appealing.

\vspace{-3.4mm}
\vspace{-1.6mm}

The main challenges for developing such a safety supervisor that unifies {\bf Safety Protection} and {\bf Safety Extension} include the following: First, in general, given an arbitrary system state, a constrained optimization problem needs to be solved to determine a control solution guaranteeing safety indefinitely \cite{ames2016control,jankovic2018robust,garone2017reference,li2021robust}. Second, when such a solution does not exist and the supervisor pursues extending safety instead, i.e., maximizing the duration of constraint violation-free operation, this involves solving a time-optimal control problem, which is inherently a difficult problem, especially in discrete time \cite{verschueren2017stabilizing}. In this paper, we develop an approach to achieving {\bf Safety Protection} and {\bf Safety Extension} (when {\bf Safety Protection} is not feasible) through solving a single optimization problem with only continuous variables. Our approach is inspired by the exponentially-weighted penalty function scheme first proposed in \cite{verschueren2017stabilizing} for minimum-time problems and extended in \cite{tang2022continuous} for DCOC problems. We enhance the scheme by a generalized penalty formulation, which leads to stronger theoretical properties compared to the formulations of \cite{verschueren2017stabilizing} and \cite{tang2022continuous}. We provide proofs for these properties.

 Therefore, the contributions of this paper are as follows:
 \begin{itemize}
     \item We develop a supervisory control scheme that unifies the abilities of {\bf Safety Protection} and {\bf Safety Extension}. This scheme minimally modifies the control input in such a way as to be able to maintain safety indefinitely when this is possible and maximizes the duration of constraint violation-free operation when not. A safety supervisor with this unified capability is new.
     \item We develop an approach that integrates the two abilities seamlessly through a single constrained optimization problem. The problem only involves continuous variables and hence can be handled with standard nonlinear programming solvers. We prove the theoretical properties of our approach. Our approach and its properties are novel.
     \item Within our general approach, we present a tailored design method for linear systems with convex constraints. Using this method, the online optimization problem reduces to a convex quadratic program with a strictly convex cost function and hence is easy to solve.
     \item We illustrate the application of the proposed safety supervisor and its properties with an automotive example.
 \end{itemize}

The organization of this paper is as follows: In Section~\ref{sec:pre}, we present the preliminaries that are useful in later sections for developing our safety supervisor. In Section~\ref{sec:safety}, we introduce the unified safety protection and extension problem that we address. In Section~\ref{sec:approach}, we develop the approach to solving the unified safety protection and extension problem through a single continuous optimization problem and prove its theoretical properties. In Section~\ref{sec:linear}, we present a tailored version of our general approach for linear systems with convex constraints. In Section~\ref{sec:example}, we illustrate our developments with an automotive adaptive cruise control example. We summarize our contributions and conclude the paper in Section~\ref{sec:conclusion}. The notations used in this paper are standard. In particular, $X \subset Y$ means that $X$ is a subset of $Y$, which allows them to be equal.

\vspace{-2mm}
\section{Preliminaries}\label{sec:pre}

We consider systems represented by the following model
\begin{equation}\label{equ:1}
x_{t+1} = f(x_t, u_t, w_t)
\end{equation}
where $x_t \in \mathbb{R}^n$ denotes the system state at the discrete-time instant $t$, $u_t \in \mathbb{R}^m$ denotes the control input at $t$, $w_t \in \mathbb{R}^p$ denotes an unmeasured disturbance input, and $f: \mathbb{R}^n \times \mathbb{R}^m \times \mathbb{R}^p \to \mathbb{R}^n$ is a twice continuously differentiable (i.e., $\mathcal{C}^2$) function. We assume that the control input must take values in a prescribed set, i.e.,
\begin{equation}\label{equ:2}
u_t \in U = \{u \in \mathbb{R}^m: h(u) \le 0\}
\end{equation}
where $h: \mathbb{R}^m \to \mathbb{R}^q$ is a $\mathcal{C}^2$ function. We also assume that the disturbance input $w_t$ takes values in a known set $W$ for all $t$, i.e., $w_t \in W$. 

Furthermore, we assume that it is desirable that the system state be maintained within a set of safe states, $X_0 \subset \mathbb{R}^n$, i.e.,
\begin{equation}\label{equ:3}
x_t \in X_0
\end{equation}
At each time $t$, in order to ensure \eqref{equ:3} holds at $t+1$, we want to select a control input $u_t$ that satisfies
\begin{equation}\label{equ:4}
f(x_t, u_t, w) \in X_0, \quad\forall w \in W
\end{equation}
or, expressed using set-theoretic notation,
\begin{equation}\label{equ:5}
f(x_t, u_t, W) \subset X_0
\end{equation}
i.e., for all possible realizations of $w_t$ in $W$, the predicted next state $x_{t+1}$ is in $X_0$. However, depending on $x_t$, such a control input $u_t$ may not exist or may not be in the set of available controls $U$. Therefore, we consider a sequence of sets, $X_k$, with $k = 1,2,\dots$, defined recursively as follows:
\begin{equation}\label{equ:6}
X_k = X_0 \cap \{x \in \mathbb{R}^n: \exists u \in U \text{ s.t. } f(x, u, W) \subset \tilde{X}_{k-1}\}
\end{equation}
where $\tilde{X}_{k-1}$ is a subset of $X_{k-1}$. The sequence of sets $X_k$ defined according to \eqref{equ:6} has the following property:

{\it Proposition 1:} If $x_0 \in X_k$, then there exist admissible state-feedback control laws $\pi_0, \dots, \pi_{k-1}$ that ensure $x_t \in X_0$ for $t = 0, \dots, k$ in spite of realizations of $w_0, \dots, w_{k-1} \in W$. Here, an admissible state-feedback control law $\pi_t$ is a map from $\mathbb{R}^n$ to $\mathbb{R}^m$ that produces a control input $u_t \in U$ given a state $x_t \in \mathbb{R}^n$.

{\it Proof:} According to \eqref{equ:6}, $x_0 \in X_k$ implies that $x_0 \in X_0$ and there exists control input $u_0 \in U$ such that $x_1 = f(x_0, u_0, w_0)$ is necessarily in $\tilde{X}_{k-1} \subset X_{k-1}$ in spite of any realization of disturbance input $w_0 \in W$. Similarly, $x_t \in X_{k - t}$ implies $x_t \in X_0$ and $\exists u_t \in U$ such that $x_{t+1} \in \tilde{X}_{k - t - 1} \subset X_{k - t - 1}$ for any $w_t \in W$, and this holds for $t = 0, \dots, k-1$. Note that the control input $u_t \in U$ above depends on $x_t$, i.e., it is a state-feedback control law. The result follows. $\blacksquare$

Proposition~1 indicates that if $x_t \in X_k$, then it is possible to ensure safety in terms of satisfying \eqref{equ:3} for at least $k$ time steps into the future even under the ``worst-case'' disturbance inputs in $W$. For Proposition~1 to hold, for each $k$, the set $\tilde{X}_{k-1}$ in \eqref{equ:6} can be chosen as an arbitrary subset of $X_{k-1}$ (or as $X_{k-1}$ itself). For later computational purposes, we assume that our choice enables the expression of the set-inclusion condition $f(x, u, W) \subset \tilde{X}_{k-1}$ as inequality constraints as follows
\begin{equation}\label{equ:7}
g_{k-1}(x, u) \le 0
\end{equation}
where $g_{k-1}: \mathbb{R}^n \times \mathbb{R}^m \to \mathbb{R}^{q_{k-1}}$ is a $\mathcal{C}^2$ function dependent on $f$, $W$, and $\tilde{X}_{k-1}$. 

{\it Remark 1:} For many systems, expressing $f(x, u, W) \subset \tilde{X}_{k-1}$ in the form of \eqref{equ:7} through a proper choice of $\tilde{X}_{k-1}$ is possible. For instance, suppose $w_t$ is additive, i.e., $f(x_t, u_t, w_t)$ can be written as $f_0(x_t, u_t) + E w_t$, where $E \in \mathbb{R}^{n \times p}$, $W$ is a closed and convex set, and $\tilde{X}_{k-1}$ is chosen to be polyhedral, i.e., can be expressed as
\begin{equation}\label{equ:8}
\tilde{X}_{k-1} = \{x \in \mathbb{R}^n: \Gamma_{k-1,j}^{\top} x \le \gamma_{k-1,j}, j = 1, \dots, q_{k-1}\},
\end{equation}
where $\Gamma_{k-1,j} \in \mathbb{R}^n$ and $\gamma_{k-1,j} \in \mathbb{R}$ for $j = 1, \dots, q_{k-1}$. Then, $f(x, u, W) \subset \tilde{X}_{k-1}$ can be expressed as \eqref{equ:7}. In particular, the $j$th row of $g_{k-1}(x, u)$ is
\begin{equation}\label{equ:9}
g_{k-1,j} (x, u) = \Gamma_{k-1,j}^{\top} f_0(x, u) + h_W(\Gamma_{k-1,j}^{\top} E) - \gamma_{k-1,j}
\end{equation}
where $h_W(\cdot)$ is the support function of $W$.

It is desirable if safety can be ensured indefinitely even in the ``worst case.'' For this,  we consider another set, $X_{\infty}$, that has the following two properties:
\begin{enumerate}
    \item $X_{\infty} \subset X_0$
    \item $x \in X_{\infty}$ implies $\exists u \in U$ such that $f(x, u, W) \subset X_{\infty}$
\end{enumerate}
The property 2) is called {\it robust controlled invariance}. Any $X_{\infty}$ satisfying the two properties yields the following result:

{\it Proposition 2:} If $x_0 \in X_{\infty}$, then there exists a time-independent admissible state-feedback control law $\pi$ that produces $u_t = \pi(x_t) \in U$ and ensures $x_t \in X_{\infty} \subset X_0$ for all $t \ge 0$ in spite of realizations of $w_t \in W$.

{\it Proof:} The result follows from properties 1) and 2). $\blacksquare$

In general, there may exist many sets that have the above two properties. For instance, $X_{\infty}$ may be defined by the sublevel set of a discrete-time barrier function~\cite{agrawal2017discrete} or relate to the previously introduced set $X_k$ as $k$ increases to infinity~\cite{li2021robust,li2022safe}. We assume that our choice of $X_{\infty}$ enables the expression of the set inclusion $f(x, u, W) \subset X_{\infty}$ as inequality constraints
\begin{equation}\label{equ:10}
g_{\infty}(x, u) \le 0
\end{equation}
where $g_{\infty}: \mathbb{R}^n \times \mathbb{R}^m \to \mathbb{R}^{q_{\infty}}$ is a $\mathcal{C}^2$ function dependent on $f$, $W$, and $X_{\infty}$.

\vspace{-2mm}
\section{Safety Protection \& Extension Problem}\label{sec:safety}

With all the sets defined in the previous section and their properties, we are now ready to introduce the main problem addressed in this paper.

At each time $t$, it is most desirable if we can select a control input $u_t \in U$ that satisfies
\begin{equation}\label{equ:11}
f(x_t, u_t, W) \subset X_{\infty} \,\,\iff\,\, g_{\infty}(x_t, u_t) \le 0
\end{equation}
because this not only ensures safety at $t+1$, i.e., $x_{t+1} \in X_0$, but also ensures recursive feasibility at $t+1$, i.e., we will be able to select a control $u_{t+1} \in U$ such that $f(x_{t+1}, u_{t+1}, W) \subset X_{\infty}$, no matter what value in $W$ the disturbance input $w_t$ takes. However, it may happen occasionally that such a control input $u_t$ does not exist, due to, for instance, occasional/intermittent action of disturbance inputs that are not within the assumed set~$W$. 

When the above situation occurs at $t$, it is reasonable to select a control input $u_t \in U$ that satisfies
\begin{equation}\label{equ:12}
f(x_t, u_t, W) \subset \tilde{X}_k \,\,\iff\,\, g_k(x_t, u_t) \le 0
\end{equation}
for $k$ as large as possible. This is because \eqref{equ:12} ensures $x_{t+1} = f(x_t, u_t, w_t) \in \tilde{X}_k \subset X_k$ for all possible values of $w_t$ in $W$, and, according to Proposition~1, we will be able to maintain safety for at least $k$ time steps into the future even under the ``worst-case'' disturbance realizations in $W$. In particular, we want to maximize $k$ because a larger $k$ corresponds to a longer duration of safe operation, and, in turn, more time to respond, to seek recovery and/or mitigate any harm.

Therefore, the problem we address can be stated as follows:

\noindent {\it Problem:} At each time $t$, select control input $u_t$ according to
\begin{equation}\label{equ:13}
u_t = \begin{cases} u_{t,\infty} & \text{ if \eqref{equ:141} is feasible} \\ u_{t,k^*} & \text{ where $k^*$ is the largest $k$ s.t. \eqref{equ:142} is feasible} \end{cases}
\end{equation}
with
\begin{subequations}
\begin{align}
& u_{t,\infty} \in \argmin_{u \in U}\, J_t(u)\, \text{ s.t. } f(x_t, u, W) \subset X_{\infty} \label{equ:141} \\
&\quad u_{t,k} \in U \text{ such that } f(x_t, u_{t,k}, W) \subset \tilde{X}_k \label{equ:142}
\end{align}
\end{subequations}
In \eqref{equ:141}, the cost function $J_t(u)$ is used to describe the quality of each feasible control input solution so that the optimal one can be selected when there are multiple feasible solutions. The subscript $t$ of $J_t(u)$ indicates that the cost function can be time-varying or dependent on other variables. In general, we assume that $J_t(u)$ is a $\mathcal{C}^2$ function for each $t$. In this paper, we consider designing a supervisory scheme that minimally adjusts a nominal control when the nominal control may lead to safety violation. For this purpose, we consider a cost function as follows
\begin{equation}\label{equ:15}
J_t(u) = \|\hat{u}_t - u\|_S^2
\end{equation}
where $\hat{u}_t$ denotes the nominal control input, $\|\cdot\|_S^2 = (\cdot)^{\top} S (\cdot)$, and $S$ is a positive-definite weighting matrix. Such a cost function penalizes the amount of adjustment to the nominal control so that the adjustment is minimized. In \eqref{equ:142}, we consider a feasibility problem and do not include a cost function because in this second case the primary goal is to maximize $k$, i.e., the duration of being able to operate without
constraint violations.

The problem \eqref{equ:13} is challenging due to the following reason: It is unknown a priori whether \eqref{equ:141} is feasible, and if not, what is the largest $k$ for which \eqref{equ:142} is feasible. In general, we need to solve \eqref{equ:141} and \eqref{equ:142} for each $k = 0,1,2,\dots$ to answer if it is feasible or not and then be able to identify the largest $k$ among the feasible ones. This will involve a tedious procedure of solving a series of optimization problems. In the next section we develop an approach to solve \eqref{equ:13} in one shot.

{\it Remark 2:} At each time $t$, depending on previous disturbance realizations and the state the system currently reaches, it is possible that a control that can extend safety beyond the previously predicted duration of safe operation or can maintain safety indefinitely (under the assumption of $w \in W$) becomes available. By repeatedly solving \eqref{equ:13} to determine control at each $t$, the proposed safety supervisor maximizes safety by steering the system state to the infinite-step safe set $X_{\infty}$ whenever possible (which can be viewed as a {\it safety recovery}) and maximizing the opportunities for safety recovery through extending the time before any actual safety violation.

\vspace{-1mm}
\section{One-Shot Optimization Approach}\label{sec:approach}
\vspace{-0.5mm}

In this section, we develop an approach to solving the problem \eqref{equ:13} in one shot. We make the following assumption:

{\it Assumption 1:} The sets $\tilde{X}_k$ and $X_{\infty}$ satisfy $X_{\infty} \subset \tilde{X}_{k+1} \subset \tilde{X}_k$ for all $k \ge 0$.

Assumption~1 is reasonable because $\tilde{X}_k$ ensures $k$-step safety and $X_{\infty}$ ensures infinite-step safety -- it is reasonable that the constraints for ensuring a longer duration of safety are stricter. For instance, it can be easily shown that if in \eqref{equ:6} $\tilde{X}_{k-1}$ is chosen to be $\tilde{X}_{k-1} = X_{k-1}$ for all $k \ge 1$, then $\tilde{X}_{k+1} \subset \tilde{X}_k$ for all $k \ge 0$. According to \eqref{equ:11} and \eqref{equ:12}, Assumption~1 also implies that if $g_{\infty}(x_t, u) \le 0$ holds, then $g_k(x_t, u) \le 0$ holds for all $k \ge 0$, and if $g_{k'}(x_t, u) \le 0$ holds for some $k' \ge 0$, then $g_k(x_t, u) \le 0$ holds for all $0 \le k \le k'$.

To develop the approach, we start by considering the following parametric optimization problem:
\begin{subequations}\label{equ:16}
\begin{align}
\min_{\,u,\, \epsilon}\quad & \hat{J}_t(u, \epsilon) \label{equ:16_a} \\
\text{s.t.}\quad & \hat{g}_{\infty}(x_t, u) \le {\bf 1} \varepsilon_{k'+1} \label{equ:16_b} \\
& \hat{g}_k(x_t, u) \le {\bf 1} \varepsilon_k, \quad k = 0,1,\dots,k' \label{equ:16_c} \\
& h(u) \le 0 \label{equ:16_d} \\
& \varepsilon_k \le \varepsilon_{k+1}, \quad k = k^*,\dots,k' \label{equ:16_e} \\
& \varepsilon_k = \eta_k, \quad k = 0,\dots,k^* \label{equ:16_f}
\end{align}
\end{subequations}
where $\hat{J}_t(u, \epsilon) = J_t(u) + \sum_{k = k^*+1}^{k'+1} \theta^{k' + 2 - k} \phi(\varepsilon_k)$; $k'$ is a pre-selected variable and represents the largest $k$ for consideration; $k^*$ is the largest $k$, $0 \le k \le k'+1$, for which \eqref{equ:141} or \eqref{equ:142} is feasible ($k^* = k'+1$ corresponds to the case where \eqref{equ:141} is feasible, and $0 \le k^* \le k'$ corresponds to the cases where \eqref{equ:141} is infeasible and \eqref{equ:142} is feasible for $0 \le k \le k^*$); $\theta > 1$ is a sufficiently large constant; ${\bf 1}$ denotes the column vector of $1$'s (of consistent dimension); $\epsilon = (\varepsilon_0, \dots, \varepsilon_{k'+1})$ are decision variables along with $u$; $\eta = (\eta_0, \dots, \eta_{k^*})$ are parameters with nominal value $\eta = 0$; $\phi(\cdot): \mathbb{R} \to \mathbb{R}$ in the cost function $\hat{J}_t$ is a $\mathcal{C}^2$ function that satisfies the following properties:
\begin{equation}\label{equ:161}
\begin{aligned}
& \phi(0) = 0, \quad \,\phi(\varepsilon) > 0 \text{ for } \varepsilon > 0, \\
& \frac{\text{d}\phi}{\text{d}\varepsilon}(0) = 1, \quad \frac{\text{d}^2 \phi}{\text{d}\varepsilon^2}(\varepsilon) \ge 0 \text{ for } \varepsilon \ge 0;
\end{aligned}
\end{equation}
and the functions $\hat{g}_k$ and $\hat{g}_{\infty}$ in the constraints \eqref{equ:16_b} and \eqref{equ:16_c} are defined according to:
\begin{enumerate}
    \item $\hat{g}_0 = g_0$;
    \item $\hat{g}_k$, for $1 \le k \le k'$, is the collection of rows of $g_k$ that are not in $\hat{g}_0, \dots, \hat{g}_{k-1}$ (two rows that are positive multiple of each other are viewed as identical);
    \item $\hat{g}_{\infty}$ is the collection of rows of $g_{\infty}$ that are not in $\hat{g}_0, \dots, \hat{g}_{k'}$.
\end{enumerate}
We make the following remarks: 1) To capture the largest $k$ for which \eqref{equ:142} is feasible, it is desirable to select $k'$ to be as large as possible. However, this will increase the number of the decision variables and constraints and hence the complexity of the problem. In practice, $k'$ can be selected based on the tradeoff between the desire to capture a larger feasible $k$ and computation complexity. 2) Many functions satisfy the properties in \eqref{equ:161} such as $\phi(\varepsilon) = \varepsilon$, $\phi(\varepsilon) = \varepsilon + a \varepsilon^b$ with any $a>0$ and $b>1$, and $\phi(\varepsilon) = e^\varepsilon - 1$. The use of this function will be elaborated in Corollary~1 and Remark~3. 3) The reason for defining the functions $\hat{g}_k$ and $\hat{g}_{\infty}$ and using them in \eqref{equ:16} instead of using $g_k$ and $g_{\infty}$ directly is to avoid repeated constraints (repeated constraints would cause a violation of constraint qualification at a minimizer where they are active and hence a violation of Assumption~2). Also, we let $\hat{g}_{k'+1} = \hat{g}_{\infty}$ to simplify notations in the following analysis.

The problem \eqref{equ:16} has the following property, which indicates that it solves the problem of interest \eqref{equ:13}:

{\it Proposition 3:} Let $z(0) = (u(0), \epsilon(0))$ be a minimizer of \eqref{equ:16} for $\eta = 0$. Then, under Assumption~1, 
\begin{enumerate}
    \item If \eqref{equ:141} is feasible or if \eqref{equ:141} is infeasible and $k'$ is selected to be greater than or equal to the largest $k$ for which \eqref{equ:142} is feasible, then $u(0)$ solves \eqref{equ:13};
    \item If \eqref{equ:141} is infeasible and $k'$ is selected to be smaller than the largest $k$ for which \eqref{equ:142} is feasible, then $u(0)$ is a solution that satisfies \eqref{equ:142} for all $0 \le k \le k'$.
\end{enumerate}

{\it Proof:} If \eqref{equ:141} is feasible, then according to the definition of $k^*$ in \eqref{equ:16}, $k^* = k'+1$ and hence \eqref{equ:16_f} implies $\epsilon(0) = \eta = 0$, i.e., $\varepsilon_k = 0$ for all $k = 0,\dots,k'+1$. Then, the cost function \eqref{equ:16_a} reduces to $\hat{J}_t(u, 0) = J_t(u)$ and the constraints \eqref{equ:16_b}-\eqref{equ:16_c} reduce to $\hat{g}_k(x_t, u) \le 0$ for all $k = 0,\dots,k'+1$. Therefore, $u(0)$ minimizes $J_t(u)$ subject to $\hat{g}_k(x_t, u) \le 0$ for all $k = 0,\dots,k'+1$ and $h(u) \le 0$. According to the definition of $\hat{g}_k$ below \eqref{equ:16}, $\hat{g}_k(x_t, u) \le 0$ for all $k = 0,\dots,k'+1$ is equivalent to $g_k(x_t, u) \le 0$ for all $k = 0,\dots,k'$ and $g_{\infty} (x_t, u) \le 0$, which, under Assumption~1, holds if and only if $g_{\infty} (x_t, u) \le 0$. Hence, $u(0)$ minimizes $J_t(u)$ subject to $g_{\infty} (x_t, u) \le 0$ and $h(u) \le 0$. Then, according to \eqref{equ:2}, \eqref{equ:11}, \eqref{equ:13}, and \eqref{equ:141}, $u(0)$ solves \eqref{equ:13}.

If \eqref{equ:141} is infeasible and $k'$ is selected to be greater than or equal to the largest $k$ for which \eqref{equ:142} is feasible, then $k^*$ in \eqref{equ:16} is defined to be the largest $k$ for which \eqref{equ:142} is feasible. In this case, \eqref{equ:16_c} and \eqref{equ:16_f} imply that $u(0)$ satisfies $\hat{g}_k(x_t, u) \le 0$ for all $k = 0,\dots,k^*$, which, according to the definition of $\hat{g}_k$ and Assumption~1, holds if and only if $g_{k^*}(x_t, u) \le 0$; and \eqref{equ:16_d} implies $h(u(0)) \le 0$. Then, according to \eqref{equ:2}, \eqref{equ:12}, \eqref{equ:13}, and \eqref{equ:142}, $u(0)$ solves \eqref{equ:13}.

The proof of case 2) is analogous to case 1). 
$\blacksquare$

Although the problem \eqref{equ:16} has the desired property above, its formulation relies on the knowledge of $k^*$, which, as discussed at the end of Section~\ref{sec:safety}, is unknown a priori. Therefore, we cannot directly use \eqref{equ:16} to solve the problem of interest \eqref{equ:13}. In what follows we introduce two additional assumptions and a related problem which does not rely on $k^*$, and we will show that we can use this related problem to solve \eqref{equ:16} and hence the problem of interest \eqref{equ:13}.

Let $z(0) = (u(0), \epsilon(0))$ denote a minimizer of \eqref{equ:16} for $\eta = 0$ and $\lambda(0)$ be its associated Lagrange multiplier vector. 

{\it Assumption 2:} The pair $(z(0), \lambda(0))$ satisfies the strong second-order sufficient conditions (SSC) for optimality.

For the SSC and checkable conditions under which the SSC are satisfied, see Theorem~2 of \cite{buskens2001sensitivity}. 

Under Assumption~2, there exist a neighbourhood $\mathcal{N} \subset \mathbb{R}^{k^*+1}$ of $\eta = 0$ and functions $z(\cdot)$ and $\lambda(\cdot)$ on $\mathcal{N}$ such that for any $\eta \in \mathcal{N}$, the pair $(z(\eta), \lambda(\eta))$ satisfies the SSC for \eqref{equ:16} and the following sensitivity result holds:
\begin{equation}\label{equ:17}
\lim_{\eta_k \to 0^+} \frac{\hat{J}_t(z(\eta^{(k)})) - \hat{J}_t(z(0))}{\eta_k - 0} = - \lambda_k(0)
\end{equation}
where $\eta^{(k)} = \eta_k e_k$, $e_k$ is the $k$th standard basis vector of $\mathbb{R}^{k^*+1}$, and $\lambda_k(0)$ is the Lagrange multiplier associated with the equality constraint $\varepsilon_k = \eta_k$ of \eqref{equ:16} for $\eta = 0$.

For the above sensitivity result and its proof, see Section~3 of \cite{buskens2001sensitivity} (esp., Theorem~3 and Equation~(29)).

{\it Assumption 3:} There exist $M,L>0$ such that
\begin{subequations}\label{equ:18}
\begin{align}
& \left|\lim_{\eta_k \to 0^+} \frac{J_t(u(\eta^{(k)})) - J_t(u(0))}{\eta_k - 0}\right| \le M \\
& \left|\lim_{\eta_k \to 0^+} \frac{\varepsilon_i(\eta^{(k)}) - \varepsilon_i(0)}{\eta_k - 0}\right| \le L
\end{align}
\end{subequations}
for all $i = k^*+1, \dots, k'+1$, $k = 0, \dots, k^*$, and $\theta>1$ sufficiently large.

Assumptions 2 and 3 lead to the following result:

{\it Lemma 1:} Let $z(0) = (u(0), \epsilon(0))$ be a minimizer of \eqref{equ:16} for $\eta = 0$. Suppose Assumptions~2 and 3 hold. Then, for $\theta>1$ sufficiently large, $z(0)$ is a minimizer of 
\begin{subequations}\label{equ:19}
\begin{align}
\min_{\,u,\, \epsilon}\quad & \hat{J}_t(u, \epsilon) + \sum_{k = 0}^{k^*} \theta^{k' + 2 - k} \phi(|\varepsilon_k|) \label{equ:19_a} \\
\text{s.t.}\quad & \hat{g}_k(x_t, u) \le {\bf 1} \varepsilon_k, \quad k = 0,1,\dots,k'+1 \label{equ:19_b} \\
& h(u) \le 0 \label{equ:19_c} \\
& \varepsilon_k \le \varepsilon_{k+1}, \quad k = k^*,\dots,k' \label{equ:19_d}
\end{align}
\end{subequations}
Recall that we let $\hat{g}_{k'+1} = \hat{g}_{\infty}$ to simplify notations. Therefore, the problem \eqref{equ:19} relates to \eqref{equ:16} but replaces the equality constraints $\varepsilon_k = \eta_k = 0$ in \eqref{equ:16_f} with penalties $\theta^{k' + 2 - k} \phi(|\varepsilon_k|)$ in the cost function.

{\it Proof:} Using \eqref{equ:17} and \eqref{equ:18}, we obtain the following bound
\begin{align}\label{equ:20}
& |\lambda_k(0)| = \left|\lim_{\eta_k \to 0^+} \frac{\hat{J}_t(z(\eta^{(k)})) - \hat{J}_t(z(0))}{\eta_k - 0}\right| \nonumber \\
&\le \left|\lim_{\eta_k \to 0^+} \frac{J_t(u(\eta^{(k)})) - J_t(u(0))}{\eta_k - 0}\right| \nonumber \\
& + \sum_{i = k^*+1}^{k'+1} \theta^{k' + 2 - i} \left|\lim_{\eta_k \to 0^+} \frac{\phi(\varepsilon_i(\eta^{(k)})) - \phi(\varepsilon_i(0))}{\eta_k - 0}\right| \\
&\le M + \sum_{i = k^*+1}^{k'+1} \theta^{k' + 2 - i} \left|\frac{\text{d}\phi}{\text{d}\varepsilon}(0)\right|\, \left|\lim_{\eta_k \to 0^+} \frac{\varepsilon_i(\eta^{(k)}) - \varepsilon_i(0)}{\eta_k - 0}\right| \nonumber \\
&\le M + \sum_{i = k^*+1}^{k'+1} \theta^{k' + 2 - i} L \le \left(\frac{M}{\theta^{k'+2-k^*}} + \frac{L}{\theta - 1}\right) \theta^{k'+2-k^*} \nonumber 
\end{align}
for $k = 0, \dots, k^*$. Then, for $\theta \ge \max(2M,2L + 1)$, we have $|\lambda_k(0)| \le \theta^{k'+2-k^*} \le \theta^{k' + 2 - k}$ for $k = 0, \dots, k^*$, i.e., the weight for $\phi(|\varepsilon_k|)$ in the cost function of \eqref{equ:19} is greater than the absolute value of the Lagrange multiplier $\lambda_k(0)$ associated with the equality constraint $\varepsilon_k = \eta_k = 0$ of \eqref{equ:16}. This implies that $\theta^{k' + 2 - k} \phi(|\varepsilon_k|)$ are exact penalties for the equality constraints $\varepsilon_k = \eta_k = 0$, $k = 0, \dots, k^*$, and the result follows from Theorem~4.6 of \cite{han1979exact}. $\blacksquare$

We now present the following optimization problem, which does not depend on $k^*$, and we will show that it can be used to solve \eqref{equ:16}:
\begin{subequations}\label{equ:21}
\begin{align}
\min_{\,u,\, \epsilon}\quad & J_t(u) + \sum_{k = 0}^{k'+1} \theta^{k' + 2 - k} \phi(\varepsilon_k)  \\
\text{s.t.}\quad & \hat{g}_{\infty}(x_t, u) \le {\bf 1} \varepsilon_{k'+1} \label{equ:21_b} \\
& \hat{g}_k(x_t, u) \le {\bf 1} \varepsilon_k, \quad k = 0,1,\dots,k' \\
& h(u) \le 0 \\
& 0 \le \varepsilon_k \le \varepsilon_{k+1}, \quad k = 0,1,\dots,k' \label{equ:21_e}
\end{align}
\end{subequations}

{\it Proposition 4:} Let $z(0) = (u(0), \epsilon(0))$ be a minimizer of \eqref{equ:16} for $\eta = 0$. Suppose Assumptions~2 and 3 hold, and \eqref{equ:19} has a unique minimizer and attains its minimum value. Then, for $\theta>1$ sufficiently large, $z(0)$ is a global minimizer of \eqref{equ:21}.

{\it Proof:} Let $Z_{\text{\eqref{equ:16}}}$, $Z_{\text{\eqref{equ:19}}}$, and $Z_{\text{\eqref{equ:21}}}$ denote the feasible regions of \eqref{equ:16} for $\eta = 0$, \eqref{equ:19}, and \eqref{equ:21}, respectively. It can be easily seen that $Z_{\text{\eqref{equ:16}}} \subset Z_{\text{\eqref{equ:21}}} \subset Z_{\text{\eqref{equ:19}}}$. Hence, if $z(0) = (u(0), \epsilon(0))$ is a minimizer (thus, a feasible point) of \eqref{equ:16} for $\eta = 0$, it is also a feasible point of \eqref{equ:19} and \eqref{equ:21}.

When Assumptions~2 and 3 hold and $\theta>1$ is sufficiently large, according to Lemma~1, $z(0)$ is also a minimizer of \eqref{equ:19}. If \eqref{equ:19} attains its minimum value and its minimizer is unique, it attains its minimum value at $z(0)$, i.e., $J_{\text{\eqref{equ:19}}}(z(0)) \le J_{\text{\eqref{equ:19}}}(z)$ for all $z \in Z_{\text{\eqref{equ:19}}}$, where $J_{\text{\eqref{equ:19}}}(z)$ denotes the cost function of~\eqref{equ:19}. On $Z_{\text{\eqref{equ:21}}}$, the cost function of \eqref{equ:21}, $J_{\text{\eqref{equ:21}}}(z)$, satisfies
\begin{align}
    & J_{\text{\eqref{equ:21}}}(z) = J_t(u) + \sum_{k = 0}^{k'+1} \theta^{k' + 2 - k} \phi(\varepsilon_k) \nonumber \\
    &= \hat{J}_t(u, \epsilon) + \sum_{k = 0}^{k^*} \theta^{k' + 2 - k} \phi(|\varepsilon_k|) = J_{\text{\eqref{equ:19}}}(z)
\end{align}
since \eqref{equ:21_e} implies $\varepsilon_k \ge 0$ for $k = 0, \dots, k^*$. Then, we have
\begin{equation}
J_{\text{\eqref{equ:21}}}(z(0)) = J_{\text{\eqref{equ:19}}}(z(0)) \le J_{\text{\eqref{equ:19}}}(z) = J_{\text{\eqref{equ:21}}}(z)
\end{equation}
for all $z \in Z_{\text{\eqref{equ:21}}}$. Thus, $z(0)$ is a global minimizer of \eqref{equ:21}. $\blacksquare$

Proposition~4 enables us to solve \eqref{equ:16}, which depends on the knowledge of $k^*$, through solving \eqref{equ:21}, which does not depend on $k^*$. In particular, after \eqref{equ:21} is solved, the value of $k^*$ can be read from the solution $z(0) = (u(0), \epsilon(0))$: $k^*$ is equal to the largest $k$ for which $\varepsilon_k = 0$. Since \eqref{equ:16} solves the original problem of interest \eqref{equ:13} according to Proposition~3, Proposition~4 also indicates that \eqref{equ:21} can be used to solve \eqref{equ:13}. The result of Proposition~4 relies on the uniqueness of minimizer of \eqref{equ:19} and requires us to identify the global minimizer of \eqref{equ:21} in order to obtain a (possibly local) minimizer of \eqref{equ:16}. The following result provides us with a method for realizing these conditions:

{\it Corollary 1:} Let $z(0) = (u(0), \epsilon(0))$ be a minimizer of \eqref{equ:16} for $\eta = 0$ that satisfies Assumptions~2 and 3 and with $\theta>1$ sufficiently large. Suppose $J_t(u)$ is strictly convex in $u$, $\phi(\varepsilon)$ is strictly convex and nondecreasing in $\varepsilon$, and $g_{\infty}$, $g_k$, $k = 0,1,\dots,k'$, and $h$ are all convex in $u$. Then, \eqref{equ:21} is a convex optimization problem with a strictly convex cost function and with $z(0)$ as its unique minimizer.

{\it Proof:} According to Lemma~2 (proof is included for completeness in Appendix), $\phi(\cdot)$ being strictly convex and nondecreasing and $|\cdot|$ being convex implies their composition $\phi(|\cdot|)$ is strictly convex. If $J_t(u)$ is strictly convex in $u$, $\phi(\varepsilon_k)$ is strictly convex in $\varepsilon_k$ for all $k = k^*+1, \dots, k'+1$, and  $\phi(|\varepsilon_k|)$ is strictly convex in $\varepsilon_k$ for all $k = 0, \dots, k^*$, the cost function of \eqref{equ:19}, $\hat{J}_t(u, \epsilon) + \sum_{k = 0}^{k^*} \theta^{k' + 2 - k} \phi(|\varepsilon_k|) = J_t(u) + \sum_{k = k^*+1}^{k'+1} \theta^{k' + 2 - k} \phi(\varepsilon_k) + \sum_{k = 0}^{k^*} \theta^{k' + 2 - k} \phi(|\varepsilon_k|)$ is strictly convex in $z = (u,\epsilon)$. Meanwhile, if $g_{\infty}$, $g_k$, $k = 0,1,\dots,k'$, and $h$ are all convex in $u$, the feasible set defined by the constraints \eqref{equ:19_b}-\eqref{equ:19_d} is a convex set of $z = (u,\epsilon)$. Therefore, \eqref{equ:19} is a convex problem with a strictly convex cost function and thus any minimizer of \eqref{equ:19} is unique and global. Since $z(0)$ is a minimizer of \eqref{equ:19} according to Lemma~1, \eqref{equ:19} has a unique minimizer and attains its minimum value (both at $z(0)$). Then, all assumptions of Proposition~4 are satisfied and, therefore, $z(0)$ is a global minimizer of \eqref{equ:21}.

Meanwhile, $J_t(u)$ being strictly convex in $u$ and $\phi(\varepsilon)$ being strictly convex in $\varepsilon$ implies the cost function of \eqref{equ:21}, $J_t(u) + \sum_{k = 0}^{k'+1} \theta^{k' + 2 - k} \phi(\varepsilon_k)$, is strictly convex in $z = (u,\epsilon)$, and $g_{\infty}$, $g_k$, $k = 0,1,\dots,k'$, and $h$ all being convex in $u$ implies the feasible set defined by the constraints \eqref{equ:21_b}-\eqref{equ:21_e} is a convex set of $z = (u,\epsilon)$. This shows \eqref{equ:21} is also a convex problem with a strictly convex cost function, and hence its minimizer $z(0)$ is not only global but also unique. $\blacksquare$

We note that our choice of $J_t(u)$ in \eqref{equ:15} is strictly convex in~$u$. A candidate for $\phi(\varepsilon)$ that satisfies the properties in \eqref{equ:161} and is strictly convex and nondecreasing is $\phi(\varepsilon) = e^\varepsilon - 1$. Another candidate for $\phi(\varepsilon)$ is $\phi(\varepsilon) = \varepsilon + a \varepsilon^2$ with a small $a>0$, which satisfies the properties in \eqref{equ:161}, is strictly convex, and is nondecreasing over the range $[-\frac{1}{2a},\infty)$. For a given problem, this second candidate can satisfy Corollary~1 locally for a sufficiently large range through a sufficiently small choice of $a>0$. Meanwhile, together with \eqref{equ:15}, it causes the cost function of \eqref{equ:21} to be a quadratic function of decision variables, which is easier to handle than that resulting from the first candidate $\phi(\varepsilon) = e^\varepsilon - 1$. Therefore, for practical applications, this second candidate is recommended.

{\it Remark 3:} In \eqref{equ:21} we use exponentially-weighted penalties for $\phi(\varepsilon_k)$, $k = 0,\dots,k'+1$, to achieve the goal of maximizing the smallest $k$ for which the condition $\hat{g}_k(x_t, u) \le 0$ is violated. A similar scheme was proposed in \cite{verschueren2017stabilizing} for minimum-time problems and later modified in \cite{tang2022continuous} for DCOC problems. But both \cite{verschueren2017stabilizing} and \cite{tang2022continuous} only considered the use of identity map $\phi(\varepsilon) = \varepsilon$ when formulating the penalties. In this paper we extend the scheme by allowing other choices of $\phi$ satisfying~\eqref{equ:161}. This extension is significant because it makes possible for \eqref{equ:19} and \eqref{equ:21} to have strictly convex cost functions by a proper choice of $\phi$ (such as $\phi(\varepsilon) = e^\varepsilon - 1$ or $\phi(\varepsilon) = \varepsilon + a \varepsilon^2$). The strict convexity can be used to prove uniqueness of minimizers of \eqref{equ:19} and \eqref{equ:21}, which is an important assumption of Proposition~4 and Corollary~1 for building up the connection between the minimizers of \eqref{equ:19} and \eqref{equ:21} and that of \eqref{equ:16}. Without this extension, i.e., using $\phi(\varepsilon) = \varepsilon$, the cost functions of \eqref{equ:19} and \eqref{equ:21} are at most convex but not strictly convex, making the unique minimizer assumption not easily checkable.

\vspace{-1mm}
\section{Design Method for Linear Systems}\label{sec:linear}

In this section, we elaborate a tailored design method for linear systems with additive disturbance inputs in the form of
\begin{equation}\label{equ:linear_Sys}
    x_{t+1} = A x_t + B u_t + E w_t    
\end{equation}
where $A$, $B$, and $E$ are matrices of consistent dimensions. The control set $U$ is assumed to be polyhedral and expressed as
\begin{equation}\label{equ:linear_U}
U = \{u \in \mathbb{R}^m: \Gamma_{u,j}^{\top} u - \gamma_{u,j} \le 0, j = 1,\dots,q \}
\end{equation}
where $\Gamma_{u,j} \in \mathbb{R}^{m}$ and $\gamma_{u,j} \in \mathbb{R}$. The disturbance inputs $w_t$ are assumed to be norm-bounded, i.e.,
\begin{equation}
W = \{w \in \mathbb{R}^p: \|w\| \le \omega\}
\end{equation}
where $\|\cdot\|$ denotes a vector norm and $\omega \ge 0$. Furthermore, we assume the set $X_0$ is polyhedral and can be expressed as
\begin{equation}\label{equ:X_0}
X_0 = \{x \in \mathbb{R}^n: \Gamma_{0,j}^{\top} x \le \gamma_{0,j}, j = 1,\dots, q_0\}
\end{equation}

Motivated by the method for designing infinite-step safe sets of linear systems proposed in \cite{li2022safe}, we design the sets $\tilde{X}_k$ as follows: We consider a virtual affine feedback law,
\begin{equation}\label{equ:linear_virtualu}
u_t = K x_t + v_t
\end{equation}
where $K \in \mathbb{R}^{m \times n}$ is a matrix such that $A_{\text{c}} = A + BK$ is Schur (all eigenvalues of $A_{\text{c}}$ are strictly inside the unit disk), and $v_t \in \mathbb{R}^m$. Then, we define $\Pi_k \subset \mathbb{R}^{n + m}$ as
\begin{align}\label{equ:Pi_k}
& \Pi_k = \Big\{(x,v): \Gamma_{0,j}^{\top} \left(A_{\text{c}}^t\, x + \Lambda_t v\right) \le \tilde{\gamma}_{0,j,t}, j = 1,\dots, q_0,  \nonumber \\
& \Gamma_{u,j}^{\top} \left(K A_{\text{c}}^t\, x + K \Lambda_t v  + v \right) \le \tilde{\gamma}_{u,j,t}, j = 1,\dots, q, t = 0, \dots, k\Big\} 
\end{align}
where $\Lambda_t = (I-A_{\text{c}}^t)(I-A_{\text{c}})^{-1}B$, $\tilde{\gamma}_{0,j,t} = \gamma_{0,j} - \sum_{\tau = 0}^{t-1} h_W(\Gamma_{0,j}^{\top} A_{\text{c}}^{\tau} E)$,  and $\tilde{\gamma}_{u,j,t} = \gamma_{u,j} - \sum_{\tau = 0}^{t-1} h_W(\Gamma_{u,j}^{\top} K A_{\text{c}}^{\tau} E)$, and we let $\tilde{X}_k = \text{proj}_x(\Pi_k)$, which is the projection of $\Pi_k$ onto the space of $x$. This method for designing $\tilde{X}_k$ yields the following result:

{\it Proposition 5:} The sets $\tilde{X}_k$ defined according to \eqref{equ:Pi_k} and $\tilde{X}_k = \text{proj}_x(\Pi_k)$ have the following properties: For each $k \ge 0$,
1) $\tilde{X}_k \subset X_k$, where $X_k$ is defined according to \eqref{equ:6} for $k \ge 1$; 2) $\tilde{X}_k$ is polyhedral and can be expressed as \eqref{equ:8}; and 3) $\tilde{X}_{k+1} \subset \tilde{X}_k$.

{\it Proof:} We start with the proof of 1). For $k = 0$, it can be easily seen that $\tilde{X}_0$ defined according to \eqref{equ:Pi_k} and $\tilde{X}_0 = \text{proj}_x(\Pi_0)$ satisfies $\tilde{X}_0 = X_0$. We now consider $k \ge 1$. For $x \in \tilde{X}_k = \text{proj}_x(\Pi_k)$, there exists $v \in \mathbb{R}^m$ such that $(x,v) \in \Pi_k$. According to the definition of $\Pi_k$ in \eqref{equ:Pi_k}, we have
\begin{align}\label{equ:P3_11}
& \Gamma_{0,j}^{\top} \big(A_{\text{c}}^t\, x + \Lambda_t v\big) = \Gamma_{0,j}^{\top} \big(A_{\text{c}}^{t-1}\, (A_{\text{c}} x + Bv) - A_{\text{c}}^{t-1} Bv + \Lambda_t v\big) \nonumber \\
&= \Gamma_{0,j}^{\top} \big(A_{\text{c}}^{t-1}\, (A_{\text{c}} x + Bv) + \Lambda_{t-1} v \big) \le \tilde{\gamma}_{0,j,t} \nonumber \\
&= \tilde{\gamma}_{0,j,t-1} - h_W(\Gamma_{0,j}^{\top} A_{\text{c}}^{t-1} E)
\end{align}
for $j = 1,\dots, q_0$ and $t = 1, \dots, k$, and
\begin{align}\label{equ:P3_12}
& \Gamma_{u,j}^{\top} \big(K A_{\text{c}}^t\, x + K \Lambda_t v  + v \big) = \Gamma_{u,j}^{\top} \big(K A_{\text{c}}^{t-1}\, (A_{\text{c}} x + Bv) \nonumber \\
& + K \Lambda_{t-1} v  + v \big) \le \tilde{\gamma}_{u,j,t-1} - h_W(\Gamma_{u,j}^{\top} K A_{\text{c}}^{t-1} E)
\end{align}
for $j = 1, \dots, q$ and $t = 1, \dots, k$. Note \eqref{equ:P3_11} and \eqref{equ:P3_12} imply
\begin{align}\label{equ:P3_13}
& \Gamma_{0,j}^{\top} \big(A_{\text{c}}^{t-1}\, (A_{\text{c}} x + Bv + Ew) + \Lambda_{t-1} v \big) \le \tilde{\gamma}_{0,j,t-1} \\
& \Gamma_{u,j}^{\top} \big(K A_{\text{c}}^{t-1}\, (A_{\text{c}} x + Bv + Ew) + K \Lambda_{t-1} v  + v \big) \le \tilde{\gamma}_{u,j,t-1} \nonumber 
\end{align}
for all $w \in W$ and $t = 1, \dots, k$, which, according to \eqref{equ:Pi_k}, implies $(A_{\text{c}} x + Bv + Ew,v) \in \Pi_{k-1}$ for all $w \in W$. Now consider $u = Kx + v$. According to \eqref{equ:Pi_k} and $A_{\text{c}} = A + BK$, $(A_{\text{c}} x + Bv + Ew,v) \in \Pi_{k-1}$ for all $w \in W$ implies $A_{\text{c}} x + Bv + Ew = A x + B(Kx + v) + Ew = Ax + Bu + Ew \in \text{proj}_x(\Pi_{k-1}) = \tilde{X}_{k-1}$ for all $w \in W$, i.e., $f(x,u,W) \subset \tilde{X}_{k-1}$; and it also implies $\Gamma_{u,j}^{\top} (K x + v) = \Gamma_{u,j}^{\top} u \le \tilde{\gamma}_{u,j,0} = \gamma_{u,j}$ for $j = 1,\dots, q$, i.e., $u \in U$. Therefore, we have shown that for any $x \in \tilde{X}_k$, there exists $u = Kx + v \in U$ such that $f(x,u,W) \subset \tilde{X}_{k-1}$, and this proves $\tilde{X}_k \subset X_k$ when $X_k$ is defined according to \eqref{equ:6}.

Then, 2) follows from that $\Pi_k$ defined according to \eqref{equ:Pi_k} is polyhedral and $\tilde{X}_k$ is the projection of $\Pi_k$ onto a subspace.

Finally, for 3), it is clear that $\Pi_k$ defined according to \eqref{equ:Pi_k} satisfies $\Pi_{k+1} \subset \Pi_k$. Therefore, since $\tilde{X}_{k+1} = \text{proj}_x(\Pi_{k+1})$ and $\tilde{X}_k = \text{proj}_x(\Pi_k)$, we have $\tilde{X}_{k+1} \subset \tilde{X}_k$. $\blacksquare$

{\it Remark 4:} The part 1) of Proposition~5 indicates that our designs of $\tilde{X}_k$ using the explicit formula \eqref{equ:Pi_k} and $\tilde{X}_k = \text{proj}_x(\Pi_k)$ are indeed subsets of $X_k$, and, therefore, satisfy Proposition~1. The part 2) means that we can use the expressions \eqref{equ:8} and \eqref{equ:9} in Remark~1 to obtain the functions $g_k$ to be used in the optimization problem \eqref{equ:16}.

For the infinite-step safe set $X_{\infty}$, we adopt the method of~\cite{li2022safe}: We define $\Pi_{\infty} \subset \mathbb{R}^{n + m}$ as
\begin{equation}\label{equ:Pi_inf}
\Pi_{\infty} = \Pi_{k'} \cap (X_0 \times \Omega')
\end{equation}
where $\Pi_{k'}$ is defined as in \eqref{equ:Pi_k} with $k'$ sufficiently large, $X_0$ is given in \eqref{equ:X_0}, and $\Omega' \in \mathbb{R}^m$ is defined as
\begin{align}\label{equ:Omega}
& \Omega' = \Big\{v: \Gamma_{0,j}^{\top} (I-A_{\text{c}})^{-1}B v \le (1-\varepsilon)\,\tilde{\gamma}_{0,j,k'}, j = 1,\dots, q_0, \nonumber \\
& \Gamma_{u,j}^{\top} \big( K (I-A_{\text{c}})^{-1} B + I \big) v \le (1-\varepsilon)\,\tilde{\gamma}_{u,j,k'}, j = 1,\dots, q \Big\} 
\end{align}
with $0 < \varepsilon \ll 1$ and $\tilde{\gamma}_{0,j,k'}$ and $\tilde{\gamma}_{u,j,k'}$ given below \eqref{equ:Pi_k}. Then, we let $X_{\infty} = \text{proj}_x(\Pi_{\infty})$. 

It can be shown that, under mild assumptions, for all $k'$ sufficiently large, the sets $\Pi_{\infty}$ and $X_{\infty}$ defined above become independent of $k'$ (see \cite{kolmanovsky1998theory,gilbert1999fast}). Furthermore, $X_{\infty}$ satisfies the desired two properties above Proposition~2 (see Proposition~3 of \cite{li2022safe}), is polyhedral (clear from \eqref{equ:Pi_inf} and \eqref{equ:Omega}), and satisfies $X_{\infty} \subset \tilde{X}_k$ for all $k \ge 0$, where $\tilde{X}_k$ is defined according to \eqref{equ:Pi_k} and $\tilde{X}_k = \text{proj}_x(\Pi_k)$.

{\it Remark 5:} The fact that $X_{\infty}$ is polyhedral means that we can use similar expressions as \eqref{equ:8} and \eqref{equ:9} in Remark~1 to obtain the function $g_{\infty}$ to be used in the optimization problem~\eqref{equ:16}. The part 3) of Proposition~5 and $X_{\infty} \subset \tilde{X}_k$ for all $k \ge 0$ verify that Assumption~1 is satisfied.

{\it Remark 6:} For linear dynamics \eqref{equ:linear_Sys}, polyhedral sets $X_{\infty}$, $\tilde{X}_k$, and $U$ (with functions $g_{\infty}$, $g_k$, and $h$ expressed as in \eqref{equ:8} and \eqref{equ:9} and in \eqref{equ:linear_U}), cost function $J_t(u)$ as in \eqref{equ:15}, and $\phi(\varepsilon)$ chosen to be $\varepsilon + a \varepsilon^2$ with sufficiently small $a>0$, the formulated problem \eqref{equ:21} is a strictly convex quadratic program, which is computationally easy.

\vspace{-1mm}
\section{Illustrative Example}\label{sec:example}

We use the following example to illustrate our proposed safety supervisor:
\begin{equation}\label{equ:ACC_Dynamics}
x_{t+1} = \begin{bmatrix} 1 & \Delta T \\ 0 & 1 \end{bmatrix} x_t + \begin{bmatrix} -\frac{\Delta T^2}{2} \\ - \Delta T \end{bmatrix} u_t + \begin{bmatrix} \frac{\Delta T^2}{2} \\ \Delta T \end{bmatrix} w_t 
\end{equation}
which represents a car-following scenario, where $x_t = (\Delta s_t, \Delta v_t)$ represent the bumper-to-bumper distance between the lead vehicle and the ego vehicle and their relative velocity, $u_t$ represents the ego vehicle's acceleration, $w_t$ represents the lead vehicle's acceleration, and $\Delta T = 0.25$ is the sampling period. The ego vehicle's acceleration $u_t$ is the control input and takes values in the set $U = [-2,2]$. The lead vehicle's acceleration $w_t$ is treated as a disturbance input and assumed to be bounded in the interval $W = [-w_{\max}, w_{\max}]$ with $w_{\max} = 1$. Our goal is to design a safety supervisor for the ego vehicle's adaptive cruise control system. We assume the following set $X_0$:
\begin{align}\label{equ:ACC_X0}
 X_0 = \left\{(\Delta s,\Delta v): 10 \le \Delta s \le 20, -5 \le \Delta v \le 5 \right\}
\end{align}
We note that we use this example for illustrating the properties of our proposed safety supervisor, and treating more elaborate modeling of car-following dynamics and constraints is not the focus of this paper.

\begin{figure}[thpb]
      \centering
      \includegraphics[width=245pt]{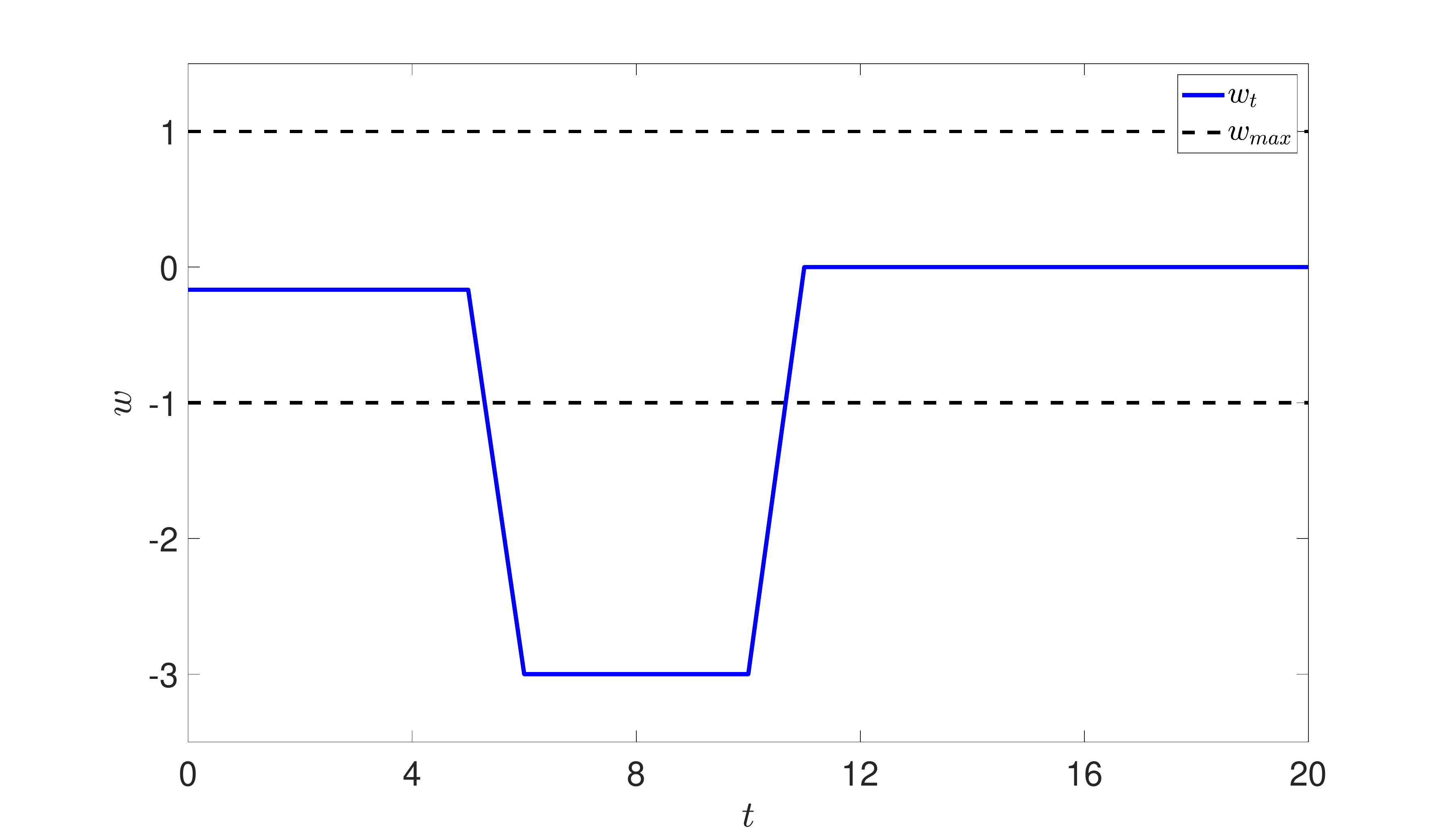}
      \caption{The disturbance input profile in Case~3.}
      \label{fig:2}
\end{figure}

We use the methods for linear systems in Section~\ref{sec:linear} to design the sets $\tilde{X}_k$, $X_{\infty}$ and functions $g_k$, $g_{\infty}$ to be used by the safety supervisor, where we use $K = [0.2842, 0.8056]$ for the virtual feedback law \eqref{equ:linear_virtualu}. For the online optimization problem~\eqref{equ:21}, we use the following parameters: $J_t(u) = 0.01 (\hat{u}_t - u)^2$, $k' = 8$, $\theta = 2$, $\phi(\varepsilon) = \varepsilon + 0.01 \varepsilon^2$. With these parameters, \eqref{equ:21} is a convex quadratic program (see Remark~6). To fully illustrate the properties of our safety supervisor, especially its unified {\bf Safety Protection} and {\bf Safety Extension} capability, we consider the following three cases:
\begin{enumerate}
    \item $x_0 = (15, 0)$ and $w_t = -w_{\max}$ for all $t \ge 0$: At this initial condition, the safety protection problem \eqref{equ:141} is feasible. The disturbance input always takes the ``worst-case'' value in $W$.
    \item $x_0 = (15, -4)$ and $w_t = -w_{\max}$ for all $t \ge 0$: At this initial condition, \eqref{equ:141} is infeasible.
    \item $x_0 = (15, 0)$ and $w_t$ follows the profile shown in Fig.~\ref{fig:2}. Over the short time period of $6 \le t \le 10$, the assumption of $w_t \in W$ is violated.
\end{enumerate}
In all three cases, we assume the nominal control: $\hat{u}_t = 0$ for all $t \ge 0$. Note that our designed safety supervisor is independent of the nominal control. Therefore, we choose $\hat{u}_t = 0$ for simplicity.

\begin{figure}[thpb]
      \centering
      \includegraphics[width=250pt]{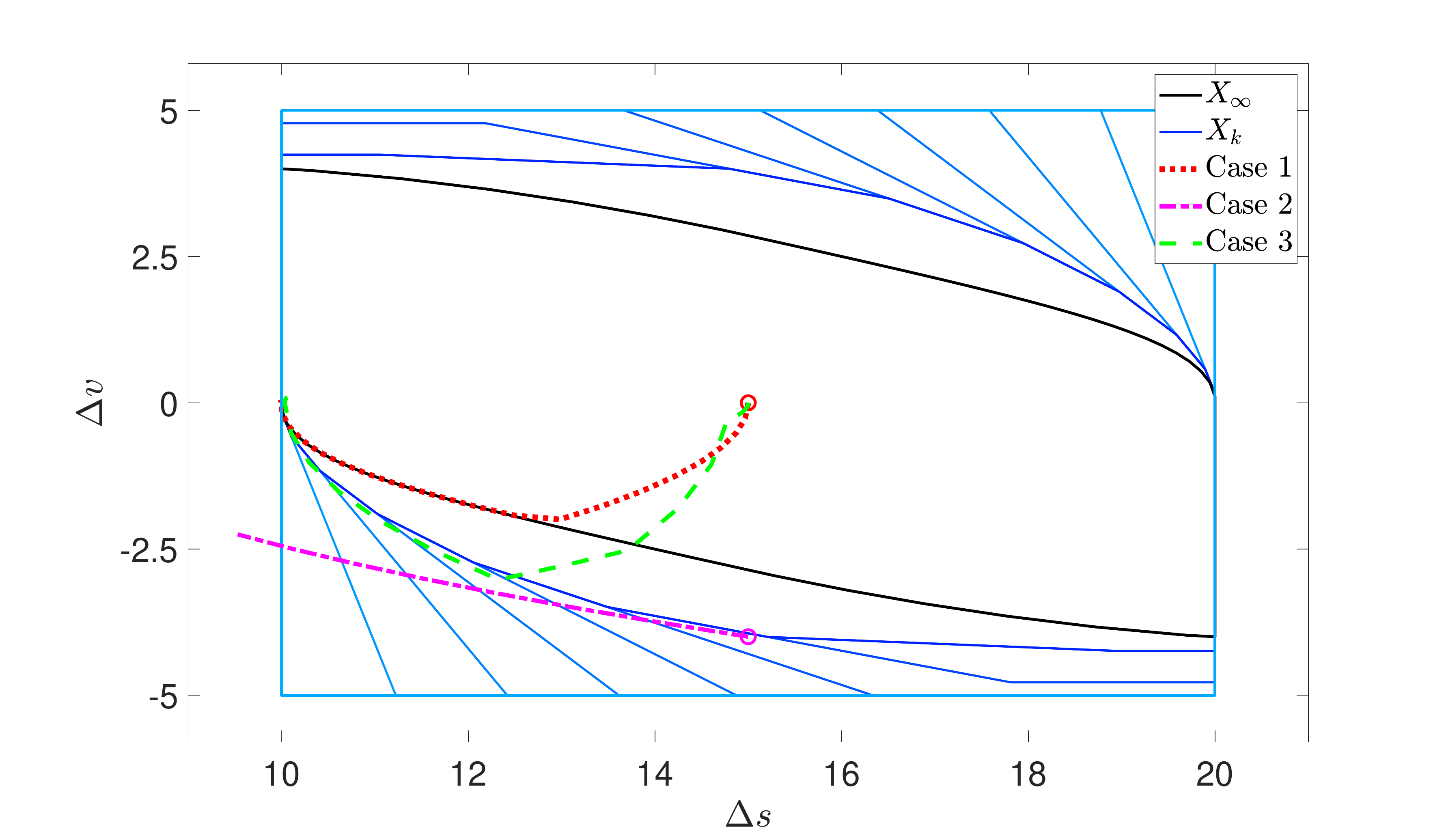}
      \caption{The state trajectories in Cases~1--3.}
      \label{fig:3}
\end{figure}

In Fig.~\ref{fig:3}, the black curve shows the boundary of the infinite-step safe set $X_{\infty}$, and the blue curves show the boundaries of the $k$-step safe sets $\tilde{X}_k$, $k = 0,\dots,k'$, where darker blue corresponds to $\tilde{X}_k$ with a larger $k$. The red dotted curve shows the trajectory of $x_t = (\Delta s_t, \Delta v_t)$ in Case~1. In Case~1, the safety protection problem \eqref{equ:141} is feasible at the initial time, and due to the robust controlled invariance property of $X_{\infty}$ and $w_t = -w_{\max} \in W$ for all $t$, \eqref{equ:141} is recursively feasible. In this case, at each $t$, the solution of our one-shot optimization problem \eqref{equ:21} is a solution to \eqref{equ:141}. Therefore, we see that our safety supervisor using \eqref{equ:21} to determine control input $u_t$ keeps $x_t$ in $X_{\infty}$ for all $t$. In particular, because the disturbance input $w_t$ is ``worst-case,'' $x_t$ reaches the boundary of $X_{\infty}$ but does not cross it. This result verifies the {\bf Safety Protection} ability of our designed supervisor.

\vspace{-4mm}
\vspace{-4mm}

\vspace{-4mm}
\vspace{-2mm}

\begin{figure}[thpb]
\begin{center}
\begin{picture}(285.0, 155.0)
\put(  -15,  -5){\epsfig{file=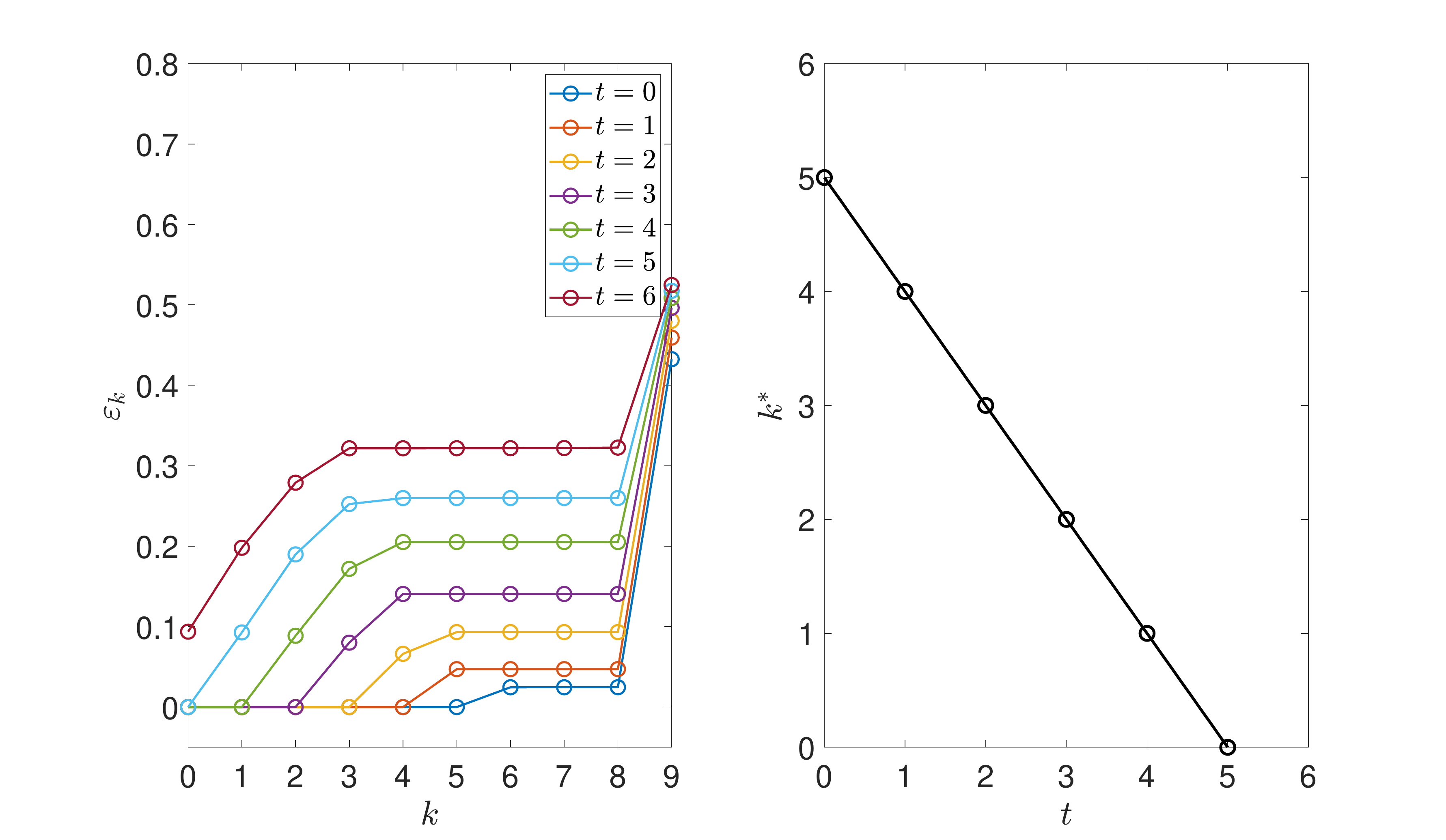,width=285pt}}
\end{picture}
\end{center}
      \caption{The $\epsilon$ and $k^*$ histories in Case~2.}
      \label{fig:4}
\end{figure}

In Case~2, \eqref{equ:141} is infeasible and our safety supervisor pursues extending safety as much as possible. For this example, it can be shown by some calculations that when the initial distance and relative velocity are $x_0 = (\Delta s_0, \Delta v_0) = (15, -4)$ and the lead vehicle brakes with a constant deceleration of $w_t = -w_{\max} = -1$, the distance constraint $\Delta s_t \ge 10$ will be violated at $t = 7$ when the ego vehicle applies the hardest brake $u_t = -2$ for all $t \ge 0$. In the numerical experiment of Case~2, our safety supervisor using \eqref{equ:21} indeed generates $u_t = -2$ at all $t \ge 0$. In Fig.~\ref{fig:4}, we show the values of $\varepsilon_k$, $k = 0,\dots,k'+1$, of the solution of \eqref{equ:21} at each $t = 0,\dots,6$ (left panel), and we also show the value of $k^*$ that is read from the values of $\varepsilon_k$ at each $t$ (right panel). Recall that $k^*$ is equal to the largest $k$ for which $\varepsilon_k = 0$. We see that $k^* = 5$ at $t = 0$, which indicates the existence of $u_0 \in U$ that ensures $x_1 \in \tilde{X}_5$ for any $w_0 \in W$. According to Proposition~1, $x_1 \in \tilde{X}_5$ implies the existence of admissible controls that ensure $x_t \in X_0$ for $t = 2,\dots,6$. Therefore, it is predicted that constraints will be satisfied until $t = 7$, and this prediction agrees with the result above derived from analytical calculations. As $t$ increases, $k^*$ decreases by one at a time. At $t = 6$, $\varepsilon_k > 0$ for all $k$ and hence $k^*$ does not exist. This means there is no $u_6 \in U$ that can enforce $x_7 \in X_0$ for all $w_6 \in W$. Correspondingly, when $w_6$ takes the ``worst-case'' value $-w_{\max}$, $x_7$ violates $X_0$. All these observations, including the generation of $u_t = -2$ (i.e., hardest brake) at all $t$ and the fact that the history of $k^*$ accurately predicts the maximum duration before the first safety violation, verifies the {\bf Safety Extension} ability of our designed supervisor.

\vspace{-2mm}
\vspace{-2mm}

In Case~3, the initial condition $x_0$ is in $X_{\infty}$. Due to the robust controlled invariance of $X_{\infty}$ and \eqref{equ:141}, $x_t$ would be in $X_{\infty}$ for all $t \ge 0$ if $w_t \in W$ for all $t$. However, due to the large disturbance input values that violate $w_t \in W$ over $6 \le t \le 10$, $x_t$ gets outside $X_{\infty}$ (shown by the green dashed curve in Fig.~\ref{fig:3}). For a control scheme that relies on the invariance of $X_{\infty}$, $x_t \notin X_{\infty}$ would cause a loss of feasibility and, consequently, a control failure (unless the constraints are relaxed). In contrast, our safety supervisor switches to {\bf Safety Extension} when $x_t \notin X_{\infty}$ occurs. By repeatedly solving \eqref{equ:21} to determine control at each $t$, our safety supervisor steers $x_t$ back to $X_{\infty}$ when disturbances are reduced, leading to a {\it safety~recovery}.

The computational advantage of our one-shot optimization approach \eqref{equ:21} is significant compared to solving \eqref{equ:13} through solving the series of optimization problems \eqref{equ:141} and \eqref{equ:142} for $k = 0,1,\dots,k'$. Taking Case~3 for example, the average computation time to determine control by solving \eqref{equ:21} is {\bf 46.5\,ms} per time step (on a {\sf Windows} laptop with {\sf i5} 1.10GHZ processor, 8GB RAM, and all implementations in the {\sf MATLAB} environment), while it takes {\bf 2.18\,s} solving the series of problems \eqref{equ:141} and \eqref{equ:142} -- our approach \eqref{equ:21} is over {\bf 40} times faster. A primary reason for the latter approach to take much longer time is that optimization solvers can consume an excessive amount of time to determine infeasibility \cite{nocedal2014interior}.

\vspace{-1mm}
\section{Conclusions}\label{sec:conclusion}

In this paper, we proposed a safety supervisor that had both {\bf Safety Protection} and {\bf Safety Extension} abilities. We developed an approach based on an exponentially-weighted penalty function scheme that integrated the two abilities in a single optimization problem with only continuous variables. For linear systems, we presented a tailored method for designing the safety supervisor such that the online optimization problem reduced to a convex quadratic program. We illustrated the proposed safety supervisor and its properties with an automotive example.

\vspace{-1mm}
\section*{}
\bibliographystyle{IEEEtran}
\bibliography{ref}

\section*{Appendix}

{\it Lemma 2:} If $f: \mathbb{R} \to \mathbb{R}$ is strictly convex and nondecreasing and $g: \mathbb{R} \to \mathbb{R}$ is convex, their composition $f \circ g: \mathbb{R} \to \mathbb{R}$ is strictly convex.

{\it Proof:} The result follows from:
\begin{align*}
& (f \circ g) \big(t x_1 + (1-t) x_2\big) = f \big(g(t x_1 + (1-t) x_2)\big) \\
&\le f \big(t g(x_1) + (1-t) g(x_2)\big) < t f\big(g(x_1)\big) + (1-t) f\big(g(x_2)\big) \\
&= t (f \circ g) (x_1) + (1-t) (f \circ g) (x_2)
\end{align*}
for all $0 \le t \le 1$ and all $x_1, x_2 \in \mathbb{R}$, where we have used the convexity of $g$ and the nondecrease of $f$ to derive the first inequality and used the strict convexity of $f$ to derive the second inequality. $\blacksquare$

\end{document}